\documentclass[%
 preprint,
superscriptaddress,
 amsmath,amssymb,
 aps,
prb,
floatfix,
]{revtex4-2}
\usepackage{setspace}
\usepackage{graphicx}
\usepackage{svg}
\usepackage{amsmath}
\usepackage{dcolumn}
\usepackage{bm}
\usepackage{physics}
\usepackage{float}
\usepackage{subcaption}
\usepackage{tabularx}
\usepackage{multirow}
\usepackage{gensymb}
\usepackage{textgreek}
\usepackage{booktabs}
\usepackage{array}

\begin{document}


\title{Anisotropic-Strain Control of The Magnetic Structure in Mn\textsubscript{3}GaN}
\author{Roman Malyshev}
\affiliation{Department of Electronic Systems, Norwegian University of Science and Technology, NO-7491 Trondheim, Norway}
\author{Ingeborg-Helene Svenum}
\affiliation{SINTEF Industry, NO-7465 Trondheim, Norway}
\affiliation{Department of Chemical Engineering, Norwegian University of Science and Technology, NO-7491 Trondheim, Norway}
\author{Sverre M. Selbach}
\affiliation{Department of Materials Science and Engineering, Norwegian University of Science and Technology, NO-7491 Trondheim, Norway}
\author{Thomas Tybell}
\affiliation{Department of Electronic Systems, Norwegian University of Science and Technology, NO-7491 Trondheim, Norway}

\date{\today}

\begin{abstract}
A first principles study is conducted to explore the changes in the magnetic structure of Mn\textsubscript{3}GaN under anisotropic biaxial strain. Mn\textsubscript{3}GaN is an antiperovskite with a structure similar to that of an ideal cubic perovskite. Several manganese nitride antiperovskites including Mn\textsubscript{3}GaN were reported to have a frustrated noncollinear antiferromagnetic structure. Successful electric switching of its magnetic structure has been reported. Furthermore, despite a cubic lattice symmetry, the magnetic symmetry is rhombohedral, allowing a piezomagnetic response. Tensile biaxial strain has been shown to produce a net magnetic moment by inducing in-plane spin canting. Compressive biaxial strain has been used to induce a spin-polarized ferro- or ferrimagnetic phase. In this study, anisotropic strain in the (001) plane is applied, outlining a magnetic phase diagram that can predict the properties when growing Mn\textsubscript{3}GaN thin films on noncubic substrates. The lattice vectors along the $a$ and $b$ crystallographic axes are strained by -5\% to 5\% in percentwise increments in all permutations. An extensive phase diagram is mapped, revealing multiple combinations of strain applied to the two lattice vectors that result in a ferro- or ferrimagnetic transition. Unlike previous results, not only strictly compressive strain on both lattice vectors, but combinations of tensile and compressive strain, as well as uniaxial strain, are seen producing the magnetic phase transitions. Furthermore, while biquadratic strain was seen producing a net moment in the $[110]$ direction under tensile strain and $[\bar{1}\bar{1}0]$ under compressive strain, anisotropic strain allows tuning the direction of the net magnetization.
\end{abstract}

\maketitle

\section{\label{sec:introduction}Introduction}
\begin{figure}[ht]
   \centering
   \begin{subfigure}{0.45\textwidth}
       \centering
       \includegraphics[width=\textwidth]{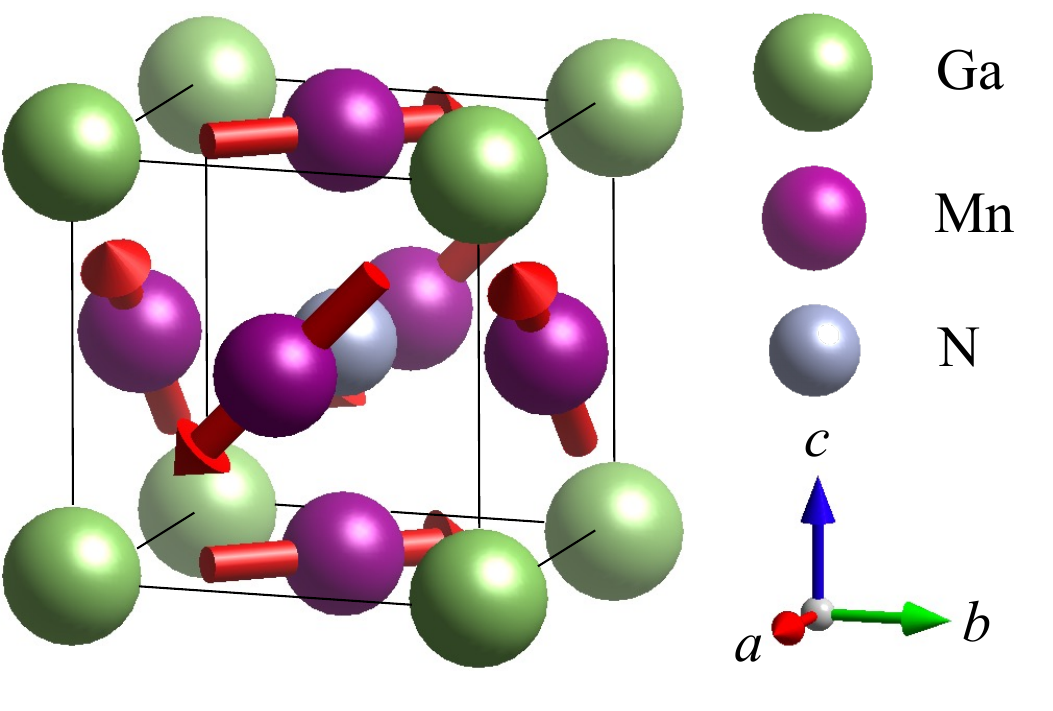}
       \caption{\label{fig:MGN_unitcell}MGN unitcell with the solid arrows illustrating the $\Gamma^\text{5g}$ spin structure.}
   \end{subfigure}
   \hfill
   \begin{subfigure}{0.45\textwidth}
       \centering
       \includegraphics[width=0.7\textwidth]{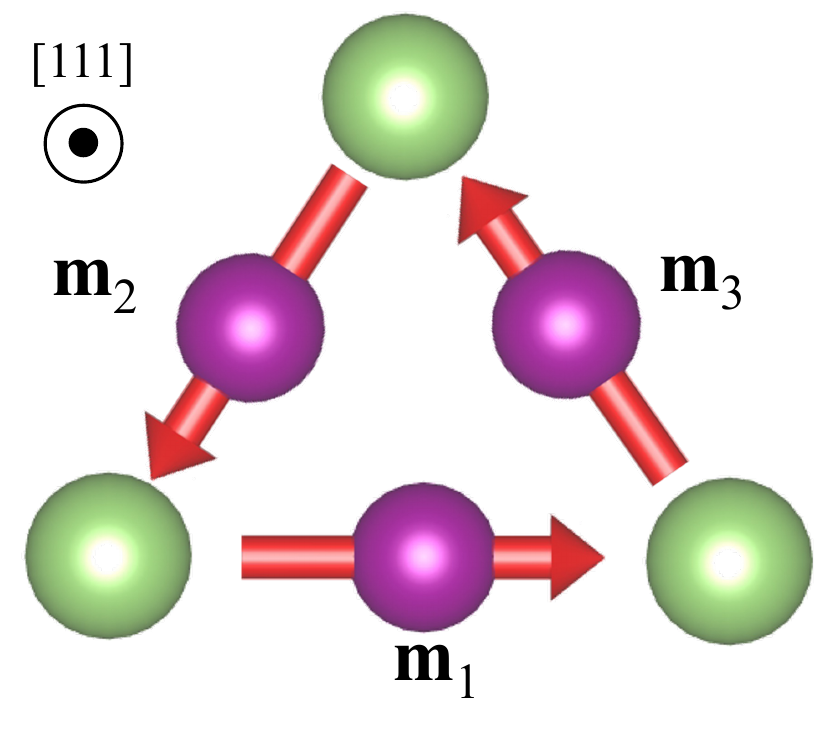}
       \caption{\label{fig:MGN_111}A view at the (111) plane, showing the noncollinear $\Gamma^{5g}$ structure in detail.}
   \end{subfigure}
    \caption{\label{fig:MGN_cell}}
\end{figure}

Antiperovskites are compounds with a structure similar to that of oxide or halide perovskites, but with the O sites occupied by cations, while the anion is found at the $B$ site. Specifically, intermetallic manganese-based antiperovskites have attracted research interest due to their demonstrated high giant magnetoresistance \cite{Mn3GaC_GMR_Kamishima_PRB, Mn3GaC_Ni_GMR_APL_2009}. Additionally, these and other classes, such as the nickel-based compounds, have demonstrated piezomagnetism, superconductivity and thermoelectricity \cite{piezomagn_Mn, piezomagn_Mn3NiN, MgCNi3_superconduct_nature_2001, CuNNi3_supercond_2013, CdCNi_supercond_2007, LaPd3P_supercond_2021, ANE_Mn3SnN, thermoel_antiperov}. Several cubic manganese nitrides -- Mn$_3$GaN, Mn$_3$ZnN, Mn$_3$AgN, Mn$_3$Cu$_{1-x}$Ge$_x$N ($x \geq 0.15$) and Mn$_3$SnN -- have been found to exhibit frustrated noncollinear antiferromagnetic (AFM) $\Gamma^{4g}$ or $\Gamma^{5g}$ (or both) orderings below their respective Néel temperatures \cite{exp_lmm_fruchart, MGN_bulk_TN, manganese_nitride_spin_polarization, fruchart_diffraction_1971, fruchart_magnetic_1978}.  Mn\textsubscript{3}GaN (MGN) has the Kagome-like $\Gamma^\text{5g}$ spin structure along the (111) plane \cite{MGN_lukashev, MGN_magnetism_caloric_effects} illustrated in Figure \ref{fig:MGN_cell}. Its magnetic unit cell is equal to the crystalline unit cell, as the propagation vector is \textbf{k} = $(0, 0, 0)$, such that the net magnetic moment per formula unit is zero. This spin structure has demonstrated, e.g., electrical spin-orbit torque switching as well as the Spin Hall effect in magnetoelectric heterostructures, making it an attractive candidate in spintronic devices, such as spin valves or magnetic memory \cite{Electrical_current_switchingMn3GaN_hajiri_ishino, mgn_sto_sciadv}.

MGN has a cubic unit cell belonging to space group \textit{Pm$\bar{3}$m}. The room temperature lattice constant in bulk material measures 3.898 \AA\, \cite{exp_lmm_fruchart}. It undergoes an antiferromagnetic-paramagnetic transition at $T_N = 345-355$ K with negative thermal compressibility. It compresses by approximately 1.1 \% cell volume \cite{MGN_neg_therm_expansion, MGN_bulk_TN, mgn_sto_sciadv}. Ferromagnetic transitions have also been explored by preparing samples with a small off-stoichiometry. A few percents N deficiency produce ferromagnetism via a tetragonal distortion in the space group $I4/mmm$, similar to that in ferromagnetic Mn$_3$Ga \cite{mn3gan1-x_ishinoa_so, ferromagnetic_mgn_lee_sukegawa_2015}. A ferrimagnetic phase named M-1 was seen at low temperatures concurrently with $\Gamma^{5g}$ by way of a tetragonal distortion by simultaneous deficiency in N and Ga, Mn$_3$Ga$_{0.95}$N$_{0.94}$ \cite{M_1_phase}. Doping with Fe ions or C was also seen reported to produce ferromagnetism without distorting the cubic phase at small concentrations of 3-4 wt\% \cite{takenaka_conversion_2009, lheritier_structures_1979}. 

The fragility of the frustrated noncollinear ordering has been reported before. Density functional theory (DFT) computations have shown that the energy difference between an AFM and FM ordering at ground state is small and can be overcome by strain. Furthermore, its magnetic symmetry is rhombohedral $R\bar3m$, not cubic as its lattice symmetry, allowing a piezomagnetic response. Thus, strain has been seen distort the spin structure, such that the angles between the spin vectors are no longer 120\degree. The in-plane spin canting produces a net magnetization. Under high levels of biaxial compressive strain the noncollinear magnetic ordering breaks down and a ferrimagnetic phase emerges \cite{MGN_magnetism_caloric_effects, anomalous_hall_effect_mgn, malyshev_effect_2025}. In the similarly-structured perovskite oxides, strain engineering has been applied to overcome the energy barriers between phases in their relatively flat energy landscape. This has allowed stabilizing otherwise energetically unfavorable functional phases \cite{adv_mat_spaldin_rondinelli_2011}. Transitions between antiferromagnetism and ferromagnetism due to epitaxial strain have been reported \cite{martinsen_amo3_2016_mrs_comm, LSMOstrainDFT2000, LSMOstrain1999, SrMnO3_multiferroic_strain}, as well as the emergence (or enhancement) of ferroelectricity \cite{sto_strained_ferroel, batio3_ferroics_strain, batio3_enhance_ferroelectric_strain, smo_strain_nature_2015}. Success has been reported growing MGN thin films on the structurally compatible cubic oxide perovskite SrTiO$_3$ \cite{nan_quintela_nature_comm, mgn_sto_2013}. These compounds have a small lattice mismatch of about 0.2\% in the plane of the SrTiO$_3$ substrate. The electronic and magnetic structure of MGN is robust to such small levels of strain. Biquadratic strain of a few percents along each crystalline axis is required to see significant changes in its electronic and magnetic structure \cite{malyshev_effect_2025}. However, the difficulty of finding substrates that can provide such levels of biquadratic strain necessitates the exploration of anisotropic strain to widen the choice of substrate.

The present density functional theory study explores biaxial anisotropic strain applied on MGN. The results predicted under biquadratic strain are expanded with a magnetic phase diagram that emerges under anisotropic strain applied to the (001) plane. It shows that not only purely compressive strain, but a combination of compressive strain along one axis and tensile along the other, or even uniaxial strain, can also induce a ferro- (FM) or ferrimagnetic (FI) transition. The symmetry of the system changes from cubic $Pm\bar3m$ to tetragonal $P4/mmm$ under all permutations of biaxial strain in the (001) plane. The electronic structure is seen changing between the magnetic phases, but is robust to strain within any given phase. The direction of magnetization in both the AFM and FM/FI phases is found to greatly depend on the combination of strain on the axes.

\section{\label{sec:methodology}Computational Method}
Density functional theory models were simulated using the \textit{Vienna Ab Initio Simulation Package} \textsc{vasp} with projected augmented wave pseudopotentials \cite{kresse96_iterative_schemes, kresse96_efficiency_total_energy, kresse_joubert_paw, blochl_paw_94}. The exchange-correlation potentials were modeled in the generalized-gradient approximation. The revised Perdew-Burke-Ernzerhof functional for solids (PBEsol) \cite{PBEsol} was used. The Mn\_pv, N and Ga\_d elemental functionals were selected, with valence states 3$p^6$4$s^2$3$d^5$, 4$s^2$3$d^{10}$4$p^1$ and 2$s^2$2$p^3$ Mn, Ga and N, respectively. The plane wave cut-off energy was set at 800 eV for optimal convergence. This makes the model computationally heavier, but allows it to converge in fewer steps and increases the accuracy, given the relatively flat energy landscape between the phases.

A $2\times2\times2$ MGN super cell was relaxed in VASP set up for noncollinear magnetism. For structural relaxations, a $4\times4\times4$ $k$ point mesh was set up. Electronic structure computations were performed with a denser $k$ mesh of $8\times8\times8$ points. The force threshold for ionic relaxation was set to EDIFFG = $-10^{-5}$ eV/\AA, while electron structure relaxations were set to stop when an energy difference of dE $= 10^{-8}$ eV was reached between the two final optimization steps. \\


Anisotropic biaxial strain was applied by stretching or compressing the supercell along the (001) plane, applying strain levels of -5\% to +5\% in one-percent increments. Strain was applied by constraining the in-plane lattice constants $a$ and $b$, but not the ion movements within the cell. The cell geometry was allowed to relax along the $c$ lattice vector. All combinations of equal (the biquadratic case, $\varepsilon_a = \varepsilon_b$) and unequal strain levels (-5\%, -5\%), (-5\%, -4\%), ..., (+5\%, +5\%) were considered according to $(\varepsilon_a, \varepsilon_b)\, | \, \varepsilon_a,\varepsilon_b \in \{-5\%, -4\%, ..., +5\% \})$. Uniaxial strain, where either $\varepsilon_a$ or $\varepsilon_b$ is zero, was also included in the set of permutations.


\section{Results and Discussion}
The relaxed structure adopts cubic $Pm\bar3m$ symmetry with a lattice constant of 3.771 \AA\, at ground state. Computational studies performed with the PBE functional \cite{PBE} reported lattice parameters of 3.860 \AA\, \cite{MGN_lukashev}, 3.864 \AA\, \cite{anomalous_hall_effect_mgn} or 3.867 \AA\, \cite{mgn_sto_sciadv}. These are closer to the experimental value 3.898 \AA\, cited in the introduction. The local magnetic moments (LMM) per Mn site were estimated at 1.91 $\mu_B$ using PBEsol. Studies based on the PBE functional estimated a value of about 2.40 $\mu_B$ \cite{MGN_lukashev, anomalous_hall_effect_mgn}. The projected atomic sphere radii are computed based on the LMMs. Thus, the smaller lattice constants achieved using PBEsol are consistent with the smaller LMMs. Recent experimental data on LMMs is lacking. The value 1.17 $\mu_B$ was reported for Mn$_3$GaN by Bertaut et al. \cite{exp_lmm_fruchart}. Their measurement of similarly structured Mn$_3$ZnN showed 1.21 $\mu_B$. More recent measurements of Mn$_3$ZnN and related compound Mn$_3$(Cu$_{1-x}$Ge$_x$)N have reported 1.8-2.2 $\mu_B$ \cite{mn3znn_kim, mn3cugen_iikubo}. Thus, the estimation of LMMs by PBEsol and PBE is within or close to this range. The Hubbard U correction was not applied. Improving the accuracy of the lattice parameter estimation using a small U correction results in inaccurate electronic structure estimation with overestimated magnetic moments compared to experimental data and PBE-based studies. This is shown in more detail in Ref. \cite{malyshev_effect_2025}, supplemental materials. \\


\begin{figure}[ht!]
    \centering
    \begin{subfigure}[t]{\linewidth}
        \centering
        \includegraphics[width=0.85\linewidth]{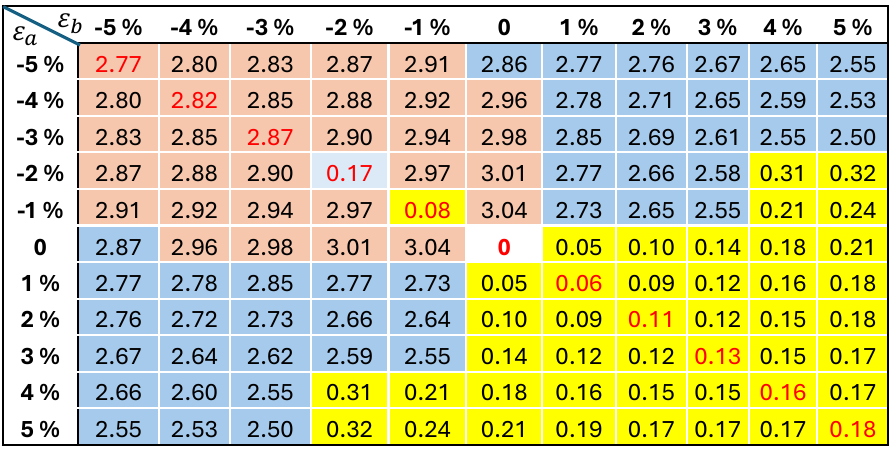}
        \caption{\label{fig:magn_moms}Net magnetic moments (in $\mu_B$) per f.u. under indicated levels $\varepsilon_a, \varepsilon_b$ of biaxial anisotropic strain applied to the $a$ and $b$ lattice vectors. The color coding indicates FM (red shading), FI (blue), $\Gamma^{5g}$ AFM (white) and distorted $\Gamma^{5g}$ (yellow).}
    \end{subfigure}
    \begin{subfigure}[t]{\linewidth}
    \centering
    \includegraphics[width=0.85\linewidth]{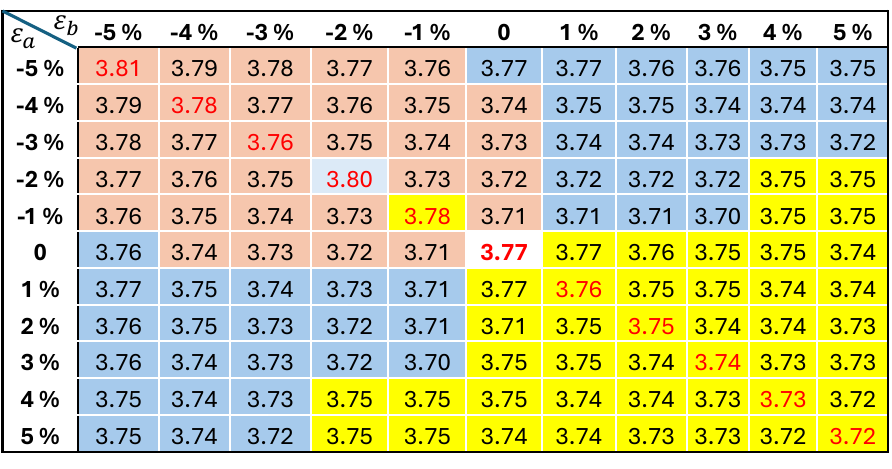}
    \caption{\label{fig:lat_consts}$c$ lattice parameter values under indicated levels of anisotropic strain levels applied to the $a, b$ lattice vectors.}
    \end{subfigure}
    \caption{\label{fig:mag_phase_diagram}}
\end{figure}

\begin{figure}
    \centering
    \includegraphics[width=0.8\linewidth]{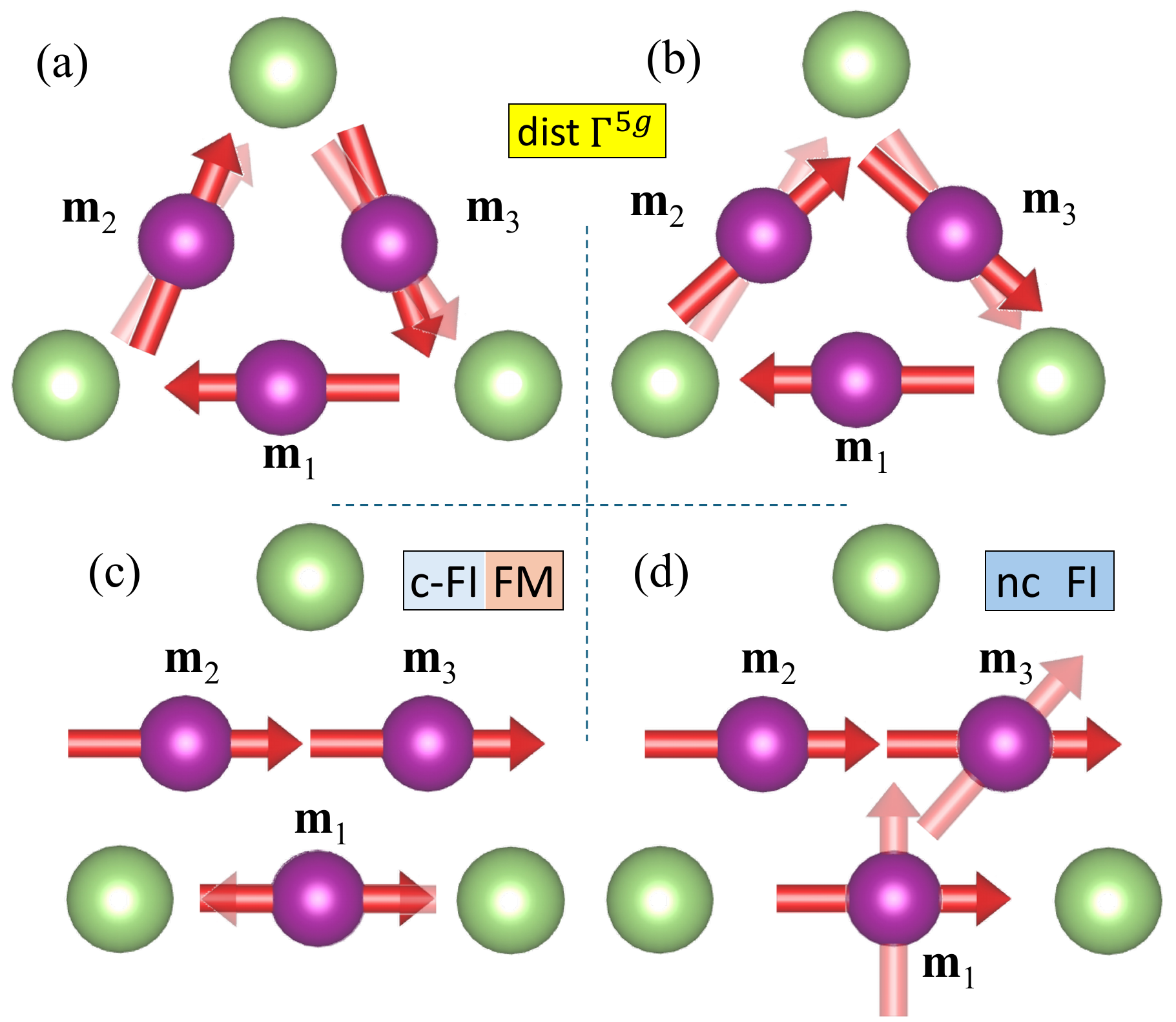}
    \caption{Magnetic structures under strain with color-coding relative to Figure \ref{fig:mag_phase_diagram}. Distorted $\Gamma^{5g}$ ordering under (a) compressive and (b) tensile strain, (c) FM ordering (also collinear FI with antiparallel $\mathbf{m}_1$) and (d) noncollinear FI orderings under anisotropic strain. Relative LMM magnitudes not shown.}
    \label{fig:magn_phases}
\end{figure}

Figure \ref{fig:magn_moms} shows the magnetic phases obtained from applying anisotropic strain to the $a$ and $b$ lattice vectors. The non-shaded central square shows the fully compensated unstrained antiferromagnetic phase seen in Figure \ref{fig:MGN_111}. The phases along the diagonal highlighted in red font emerge under the application of equal (biquadratic) strain along both $a$ and $b$ lattice vectors. Along this diagonal, the compensated $\Gamma^{5g}$ magnetic ordering becomes distorted. A net moment emerges under both compressive and tensile strain as the cubic lattice undergoes a tetragonal distortion, shown in figures \ref{fig:magn_phases}a and \ref{fig:magn_phases}b, respectively. This is consistent with previous studies \cite{malyshev_effect_2025, zemen_frustrated_2017, MGN_lukashev}. Tensile strain preserves the triangular magnetic ordering, but canting of spins is introduced by the lattice distortion. All non-zero strain levels that preserve the noncollinear $\Gamma^{5g}$-like structure, but distort it, induce a net moment and are color-coded in yellow. At -2\% compressive strain, an intermediate ferrimagnetic (FI) phase with antiparallel unequal spins on the three Mn sites emerges and is color-coded in light blue. The spins flip under higher compressive strain and are parallel at -3\%. The local magnetic moments (LMM) are unequal in this phase as well, such that the phase resembles ferromagnetism, but with up to three distinct sublattices, one of which is unequal to the other two. The unequal spin is located on the apex Mn site, while the two equatorial sites retain equal spins. This parallel ferrimagnetic phase is labeled FM due to the likeness to ferromagnetism and color-coded with a red shading. Figure \ref{fig:magn_phases}c shows the spin ordering in these collinear FI and FM phases. Along the diagonal, the directions of the net magnetic moment are [$\bar1 \bar1 0$] and [110] for compressive and tensile strain, respectively.

The direction of magnetization under anisotropic strain depends on the level of strain along each axis. The magnetization axis changes with the combination of strain on the $a$ and $b$ axes.
Furthermore, under the application of off-diagonal \textit{anisotropic} strain the noncollinear $\Gamma^{5g}$ ordering is less stable. Uniaxial compressive strain of -1\% is enough to transform the magnetic ordering into a FM phase with parallel spins. Furthermore, a new noncollinear ferrimagnetic state emerges under uniaxial tensile strain and highly anisotropic biaxial strain, where one lattice parameter is compressed while the other is under tensile strain. The onset of this ordering is seen at ($\varepsilon_a, \varepsilon_b$) = (1\%, -1\%). At least one of the LMMs in this phase is canted in relation to the other spin vectors. The spin ordering is shown in Figure \ref{fig:magn_phases}d. In the limiting case where $\mathbf{m}_1$ points inwards at a normal angle in relation to the two other spins, this is similar to the M-1 FI phase described in the introduction. However, the canting of the local moments depends on the strain level ($\varepsilon_a, \varepsilon_b$). In the tensile strain quadrant of Figure \ref{fig:magn_moms}, the $\Gamma^{5g}$ ordering is preserved, albeit distorted. Some combinations of high tensile strain along one axis, 4-5\%, and compressive strain of 1-2\% along the second axis also this ordering. 

From Figure \ref{fig:magn_moms} it is seen that all non-zero strain levels induce a net moment. The tabulated values of net magnetic moments are specified per formula unit. The net moments in the distorted $\Gamma^{5g}$ phase, yellow fields, are all on the order of a few tenths of a $\mu_B$, producing weak ferromagnetism. The emergence of these net moments is due to canting, as seen in figures \ref{fig:magn_phases}a and \ref{fig:magn_phases}b, which increases with strain. The lightly red-shaded FM phases show the highest moments, as expected, due to all spins being parallel. The small variations are also due to slight canting of spins on the order of 1.0\degree\, relative to one another. Uniaxial (-1\%, 0) or (0, -1\%) strain shows the maximum degree of spin polarization. This means that switching the magnetic structure from fully compensated AFM to FM does not require compressive biquadratic strain levels of at least -3\%, as indicated by the red-font diagonal, but can be achieved with a much lower lattice mismatch that does not need to be symmetric between the axes. 

The noncollinear FI phase emerging under anisotropic strain (blue shading) also achieves net moments of a similar magnitude to that of the FM phase. However, in this phase, at least one of the LMMs is significantly canted relative to the magnetization axis, which does not allow producing the maximal values seen in the FM phase. Together with the FM phase, the strongly spin-polarized phases make up three quadrants of the table in Figure \ref{fig:magn_moms}, showing the many possible ways to induce an AFM-FI/FM magnetic transition in MGN. 

The magnetic transitions are consistent with first-order phase transitions, exhibiting a discontinuity in cell volume and energy. This is illustrated by the tabulated values of the relaxed $c$ lattice parameter in Figure \ref{fig:lat_consts}. The AFM-FM and AFM-FI transitions show volume contractions, similar to the NTE observed during the AFM-PM transition at the Néel temperature. This compares well with stoichiometry- or doping-induced AFM-FM or AFM-FI transitions reported at low temperatures \cite{mn3gan1-x_ishinoa_so, ferromagnetic_mgn_lee_sukegawa_2015, M_1_phase, takenaka_conversion_2009, lheritier_structures_1979}.

\section{Conclusion}
Anisotropic biaxial strain is applied to the lattice vectors of the MGN supercell in the (001) plane in order to explore its implications for the magnetic and electronic structure. While applying biquadratic strain results in a FM or FI transition only at high levels of compression, anisotropic strain in the (001) plane gives many more possibilities to tailor the lattice mismatch to the desired magnetic phase. FI phases can thus be produced when strain along only one axis is compressive, while the strain along the other in-plane axis is tensile. Furthermore, uniaxial strain is predicted to produce direct AFM-FM transitions that yield the highest net magnetization. This allows for many more possible substrates to exert control on the magnetic structure when MGN is grown as a thin film or in heterostructures. Under biquadratic strain with equal compression of $a$ and $b$ lattice vectors, the spin vectors all point in the $[\bar{1}\bar{1}0]$ direction. Under tensile strain with equal elongation of both lattice vectors the net magnetization points in $[110]$. Anisotropic strain allows tuning the magnetization vector further.

\section{Acknowledgments}
The Norwegian Research Infrastructure Services are acknowledged for providing computational resources at
Uninett Sigma 2, project No. NN9301K.

\section{Data Availability Statement}
The code and data used in the preparation of this manuscript are publicly available \cite{data}.
\bibliography{apssamp}

\end{document}